\documentclass[twocolumn]{aastex701}
\usepackage{newtxtext,newtxmath}
\usepackage[T1]{fontenc}
\usepackage{graphicx}	
\usepackage{amsmath}	
\usepackage{amssymb}	
\usepackage{tabularx}

\usepackage{xcolor}
\usepackage{hyperref}
\usepackage{subfigure}
\usepackage{soul}
\usepackage{siunitx}
\usepackage{twoopt}

\usepackage{totcount}

\newtotcounter{citnum} 
\def\oldbibitem{} \let\oldbibitem=\bibitem
\def\bibitem{\stepcounter{citnum}\oldbibitem}

\shorttitle{MaDCoWS2: splashback radii in total light stacks}
\shortauthors{Trudeau et al.}

\graphicspath{{./}{figures/}}

\begin{document}
\defcitealias{trudeau_massive_2024}{T24}
\defcitealias{murata_splashback_2020}{M20}
\defcitealias{thongkham_massive_2026}{T26}

\title{The Massive and Distant Clusters of WISE Survey 2: Detection of splashback radii in galaxy cluster total light stacks}

\author[0000-0003-3428-1106]{A. Trudeau}
\email{atrudeau@asiaa.sinica.edu.tw}
\affiliation{Institute of Astronomy and Astrophysics, Academia Sinica, Taipei 10617, Taiwan}
\affiliation{Department of Astronomy, University of Florida, 211 Bryant Space Center, Gainesville, FL 32611, USA}

\author[0000-0002-0933-8601]{Anthony H. Gonzalez}
\email{anthonyhg@astro.ufl.edu}
\affiliation{Department of Astronomy, University of Florida, 211 Bryant Space Center, Gainesville, FL 32611, USA}

\author[0000-0001-7027-2202]{K. Thongkham}
\email{khunanon@narit.or.th}
\affiliation{National Astronomical Research Institute of Thailand (NARIT), Mae Rim, Chiang Mai, 50180, Thailand}
\affiliation{Korea Astronomy and Space Science Institute, 776, Daedeokdae-ro, Yuseong-gu, Daejeon 34055, Republic of Korea}
\affiliation{Department of Astronomy, University of Florida, 211 Bryant Space Center, Gainesville, FL 32611, USA}

\author[0000-0002-4208-798X]{M. Brodwin}
\email{brodwin@eurekasci.com}
\affiliation{Eureka Scientific, 2452 Delmer Street Suite 100, Oakland, CA, 94602-3017, USA}

\author[0000-0002-7898-7664]{Thomas Connor}
\email{thomas.connor@cfa.harvard.edu}
\affiliation{Center for Astrophysics $\vert$\ Harvard\ \&\ Smithsonian, 60 Garden St., Cambridge, MA 02138, USA}

\author{Peter R. M. Eisenhardt}
\email{Peter.R.Eisenhardt@jpl.nasa.gov}
\affiliation{Jet Propulsion Laboratory, California Institute of Technology, 4800 Oak Grove Dr., Pasadena, CA 91109, USA}

\author[0000-0001-9793-5416]{Emily Moravec}
\email{emoravec@nrao.edu}
\affiliation{Green Bank Observatory, P.O. Box 2, Green Bank, WV 24944}

\author[0000-0003-0122-0841]{S. A. Stanford}
\email{stanford@physics.ucdavis.edu}
\affiliation{Department of Physics, University of California, One Shields Avenue, Davis, CA, 95616, USA}

\author[0000-0003-2686-9241]{D. Stern}
\email{daniel.k.stern@jpl.nasa.gov}
\affiliation{Jet Propulsion Laboratory, California Institute of Technology, 4800 Oak Grove Dr., Pasadena, CA 91109, USA}

\begin{abstract}

The splashback radius, the radius of the apocenter of the first orbit of infalling material, is a measurable quantity marking the boundary between a galaxy cluster and its infalling region. We report detections of splashback radii in total light stacks, i.e. image stacks centered on the cores of galaxy clusters. Our analysis uses Wide-field Infrared Survey Explorer (WISE) W1 and W2 images of 83,345 candidate clusters at $0.5 \lesssim z \lesssim 1.9$ from the Massive and Distant Clusters of WISE Survey 2 (MaDCoWS2). The clusters are organized in stacks by redshift and signal-to-noise ($S\slash N$) ratios. We adopt a statistical approach, using 1000 bootstrap realizations to determine the median projected splashback radius and its confidence interval in a given bin. We compare our splashback radii with the measurements made by K. Thongkham et al. on a similar sample of MaDCoWS2 clusters using galaxy-cluster cross-correlation and find that they are consistent, although our method yields larger error bars. Our main systematic error is the accuracy of the background subtraction, but its impact remains small: the consistency of K. Thongkham et al. and our results suggests that neither method suffers from large systematics. The sensitivity of total light stacking to the contribution of faint galaxies can be advantageous to locate splashback radii when only the brightest galaxies are detected in individual images, such as at high redshifts. We present a potential application of this new technique to probe the evolution of the stellar mass in cluster infalling regions.

\end{abstract} 

\keywords{Cosmic web (330) --- Galaxy clusters (584) --- Galaxy environments (2029) --- Galaxy evolution (594) --- High-redshift galaxy clusters (2007) --- Infrared astronomy (786)}


\section{Introduction} \label{sec_intro}

Clusters of galaxies sit at the nodes of the cosmic web \citep{zeldovich_gravitational_1970,bond_how_1996} and grow by accreting other structures \citep[e.g.][]{gunn_infall_1972,behroozi_gravitationally_2013,wu_rhapsody_2013,muldrew_what_2015,klypin_multidark_2016}. Their progenitors are loosely bound, unvirialized collections of halos called protoclusters \citep[e.g.][]{chiang_ancient_2013,muldrew_what_2015,lovell_characterising_2018}.

The transition from a protocluster to a cluster is usually defined as the point at which a collapsed core with a mass exceeding $10^{14}~M_\odot$ is formed \citep{chiang_ancient_2013,muldrew_what_2015,lovell_characterising_2018,remus_young_2023}, though these clusters remain embedded in a larger scale overdensity. Observationally, clusters are distinguished from protoclusters by two main observables: the presence of a hot intracluster medium and/or the buildup of a red-sequence \citep[e.g.][]{stanford_evolution_1998,papovich_spitzer-selected_2010,gobat_mature_2011,brodwin_idcs_2012,brodwin_idcs_2016,andreon_jkcs_2014,mantz_xxl_2014,mantz_xxl_2018,willis_spectroscopic_2020}.

By definition, protoclusters are diffuse structures, extending over 10-40 comoving Mpc \citep{steidel_large_1998,suwa_protoclusters_2006,chiang_ancient_2013,chiang_discovery_2014,chiang_galaxy_2017,muldrew_what_2015} although several authors claim to have found larger protoclusters \citep[e.g.][]{hung_large-scale_2016,cucciati_progeny_2018,shi_how_2019,wang_100_2025}. The size of a protocluster is difficult to gauge:
as pointed out by \citet{overzier_realm_2016} and \citet{lovell_characterising_2018}, protoclusters are not clearly divided from the rest of the cosmic web.

For collapsed clusters, the most commonly used size indicator is the radius enclosing a density equivalent to a multiple of the critical density, typically 200 or 500. Albeit less arbitrary than protocluster sizes, these radii, denoted $r_{200}$ and $r_{500}$, do not correspond to physically-motivated boundaries \citep[e.g.][]{more_splashback_2015,deason_stellar_2021}. They also depend on the redshift and assumed cosmology \citep{tinker_toward_2008,watson_halo_2013,diemer_universal_2020}. 

When material falls into a halo, the apocenter of its first orbit \citep{gunn_infall_1972,fillmore_self-similar_1984,bertschinger_self-similar_1985}
appears as a sharp break or a slope steepening in the halo dark matter profile \citep[e.g.][]{adhikari_splashback_2014,diemer_dependence_2014,more_splashback_2015,diemer_dynamics-based_2023}. As the separation between infalling and collapsed material, this ``splashback radius'' has been proposed as a more physically-motivated halo boundary \citep{more_splashback_2015}. Further numerical efforts showed that while splashback radii depend weakly on redshift and cosmology, they are mostly sensitive to accretion rates \citep{adhikari_splashback_2014,diemer_splashback_2017-1,diemer_universal_2020,deason_stellar_2021, shin_what_2023}.

Observationally, splashback radii have been detected in weak-lensing cluster profiles \citep[e.g.][]{umetsu_lensing_2017,chang_splashback_2018,shin_measurement_2019,shin_mass_2021,contigiani_weak_2019,fong_first_2022,giocoli_amico_2024}, or through tracers of the dark matter distribution. These tracers include galaxy counts \citep[usually estimated through galaxy—cluster cross-correlations, e.g.][see also \citealt{murata_splashback_2020} and \citealt{thongkham_massive_2026}, hereafter \citetalias{murata_splashback_2020} and \citetalias{thongkham_massive_2026}]{more_detection_2016,baxter_halo_2017,shin_mass_2021,rana_erosita_2023} and intracluster light \citep[e.g.][]{gonzalez_discovery_2021}. Sunyaev-Zel'dovich profiles also contain shocks at the general location of the splashback radii \citep{anbajagane_shocks_2022,anbajagane_cosmological_2024,lebeau_can_2024,towler_inferring_2024,zhang_three_2025}.

Splashback radii traced by galaxy counts yield systematically smaller radii than predictions based on simulated dark matter haloes \citep[e.g.][\citetalias{murata_splashback_2020}]{more_detection_2016,xhakaj_how_2020,oneil_splashback_2021,oshea_dynamical_2025}. However, there are differences between galaxy populations: the splashback radii traced by low-mass red galaxies are closer to the dark matter values than the splashback radii of massive red \citep{dacunha_connecting_2022,oneil_impact_2022} or blue galaxies \citep[][\citetalias{murata_splashback_2020}]{adhikari_probing_2021,dacunha_connecting_2022,oneil_impact_2022}. The larger dynamical friction experienced by massive galaxies is the most common explanation for their smaller splashback radii \citep[e.g.][]{adhikari_observing_2016,xhakaj_how_2020,dacunha_connecting_2022,oneil_impact_2022,oshea_dynamical_2025}, though \citet{adhikari_probing_2021} and \citet{dacunha_connecting_2022}, point out that blue galaxies might also have accreted too recently to reach their first apocenter.

Most of the observational works cited above focus on $z \lesssim 0.7$ since measuring high-redshift splashback radii is challenging. One of the first studies to probe into earlier epochs is \citetalias{thongkham_massive_2026} which use galaxy-cluster cross-correlations to determine splashback radii out to $z\sim 1.5$ for the Massive and Distant Clusters of the WISE Survey (MaDCoWS2) sample. Their measurements are very precise because of the large size of their sample — 133,036 clusters in total, with 24,782 clusters at $z\gtrsim 1$.

We originally intended to measure the average stellar mass of MaDCoWS2 clusters, but reoriented our research upon serendipitously detecting splashback radii in our total light stacks. Thus, in this paper we present the detection of projected splashback radii in stacked near-infrared images of MaDCoWS2 galaxy clusters. The independent measurements by \citetalias{thongkham_massive_2026} provide a robust comparison sample to assess the performance of this new method.

Common splashback radii detection techniques usually involve ensemble averaging: either by generating composite weak-lensing profiles \citep[e.g.][]{giocoli_amico_2024} of by calculating galaxy cluster cross-correlations \citepalias[e.g.][]{thongkham_massive_2026}. Our approach of stacking imaging data also involves ensemble averaging, but represents a technique that has not previously been employed for this scientific application. For this work, we will specifically be using weighted mean stacks of imaging centered on the galaxy cluster positions. Weighted mean imaging stacks have previously in literature also been referred to as ``total light stacks,'' \citet{alberts_measuring_2021}, a terminology that we adopt for the rest of this paper.

Our technique is a variation over the more conventional image stacks centered on member galaxies. Indeed, \citet{bianconi_locuss_2021}, who claim the first detection of a splashback radius in a stacked cluster luminosity profile, use the mean K-band magnitudes of spectroscopic cluster members. The main advantage of total light stacks is the sensitivity to the statistical contribution of all components emitting at the considered wavelengths. Near-infrared stacks of galaxy clusters will thus take into account the contributions of dwarf galaxies and the intracluster light \citep[ICL; e.g.][]{kelly_60_1990,zibetti_intergalactic_2005,alberts_measuring_2021,popescu_tracing_2023}, while galaxy-cluster cross-correlation techniques and \citet{bianconi_locuss_2021} luminosity stacks are sensitive only to the galaxies bright enough to be detected individually. Both the fainter galaxy \citep{dacunha_connecting_2022,oneil_impact_2022} and the ICL \citep{deason_stellar_2021,gonzalez_discovery_2021} profiles will display a drop at the location of the splashback radius.

This article is divided as follow: Section \ref{sec_data} briefly presents the cluster sample, the Wide-field Infrared Survey Explorer (WISE) images \citep{wright_wide-field_2010} and the total light stacking technique. Section \ref{sec_results} presents the splashback radii determination and modeling, while Section \ref{sec_discussion} compares our radii to those obtained for the same sample by \citetalias{thongkham_massive_2026} and discusses systematics. A possible application of our method is presented in Section \ref{sec_infalling}, and our main conclusions are summarized in Section \ref{sec_conclusion}. Throughout this paper, we use AB magnitudes and assume a \citet{chabrier_galactic_2003} initial mass function. All scales are in physical, non-comoving units unless otherwise stated. We assume the \citet{collaboration_planck_2020} $\Lambda$CDM cosmology as implemented in {\tt astropy.cosmology}: $\Omega_m =0.31$ and $H_0 = 67.7$ km s$^{-1}$ Mpc$^{-1}$.

\section{Data, processing, and stacks}\label{sec_data}

\subsection{Cluster catalog and sample}\label{ssec_madcows2}

MaDCoWS2 is a galaxy cluster survey based on $\mathrm{grz}$ bands from the Dark Energy Camera (DECam) Legacy Survey \citep[DECaLS; see][]{flaugher_dark_2015,dey_overview_2019} and on the W1 and W2 data from the CatWISE2020 catalog \citep{eisenhardt_catwise_2020,marocco_catwise2020_2021}. The second data release of MaDCoWS2 \citep{thongkham_massive_2024b} includes the first data release \citep[$1461\deg^2$, see][]{thongkham_massive_2024a} and covers a total of $6498\deg^2$. We refer the reader to these two works for details\footnote{The catalog is available at \url{https://doi.org/10.26131/irsa583}}.

The $\mathrm{grzW1W2}$ photometry is used to compute photometric redshift probability distribution functions \citep[PDFs; see][]{brodwin_photometric_2006} that are then ingested into the PZWav algorithm \citep[e.g.][]{euclid_collaboration_euclid_2019,werner_s-plus_2023}. PZWav creates a series of density maps at $0.1\leq z\leq 3.0$ in increments of $\Delta z = 0.06$. Each map has a pixel size of 12 arcsec and is convolved with a difference-of-Gaussians kernel ($\sigma=0.4$ and 2 Mpc for the inner and outer Gaussian, respectively). The resulting candidate clusters are classified by their signal-to-noise ratios ($S\slash N$), which correlate with the mass and redshift. A convolutional neural network (CNN) estimates the probability that $z\geq 1$ candidates are real rather than due to artifacts, which otherwise increase contamination at the highest redshifts.

The MaDCoWS2 public catalog contains every cluster detection with $S\slash N \geq 5$ and $z\leq 2$. In this work, we restrict our sample to $0.5\leq z \leq 1.92$ candidates with CNN probabilities $>0.5$ for the redshifts at which this probability is defined. We choose to cut our sample at $z=1.92$ rather than $z=2$ to accommodate our recursive redshift binning. These cuts result in a sample of 83,345 candidate clusters.

\begin{figure}
\centering
\includegraphics[width=\columnwidth]{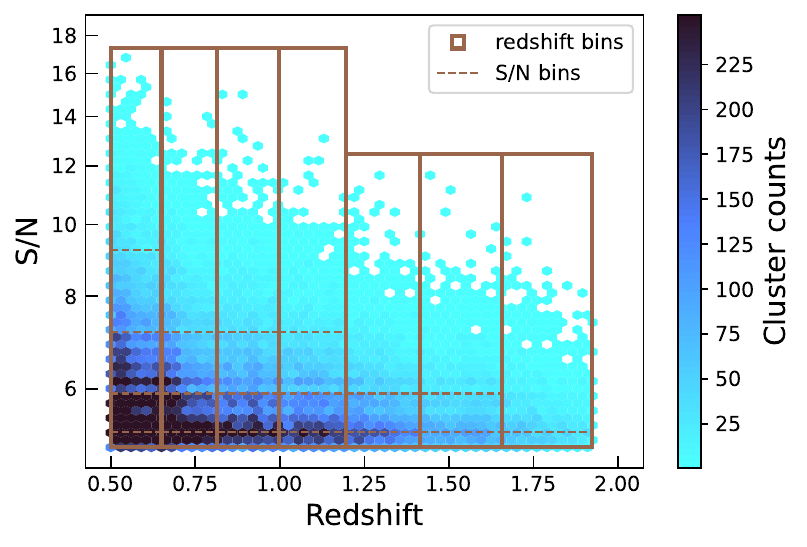}
\caption{The photometric redshift and $S\slash N$ subdivisions used in this work. The cluster sample is represented by hexagons, color-coded by counts. The main redshift divisions are indicated as rectangles delimited by full lines, with the height of the rectangle giving the maximum S/N in a redshift bin. The dashed lines show the $S\slash N$ subdivisions, with values listed in Table \ref{tab_binning}.}
\label{fig_binning}
\end{figure}

To account for the increased uncertainty of the photometric redshifts at higher redshift, we use recursive redshift binning as in \citet[][hereafter \citetalias{trudeau_massive_2024}]{trudeau_massive_2024}: $\Delta z_i = 0.1 (1+z_{min,i})$, where $z_{min,i}$ is the lower limit of a given photometric redshift bin. Our $S\slash N$ increments are based on $\Delta S\slash N_i = -0.75 +10^{0.15i}$, with a minimal $S\slash N=5$. Hence, $\Delta S\slash N_0=0.25$, $\Delta S\slash N_1=0.66$, $\Delta S\slash N_2=1.25$, and $\Delta S\slash N_3=2.07$. Additionally, each $S\slash N$ subdivision must contain at least 750 clusters. If it does not, it is merged with lower $S\slash N$ subdivisions until the combined cluster count exceeds 750. This binning corresponds to a compromise between maintaining a high and relatively uniform number of clusters per bin, given the increasingly rare higher mass halos \citep[e.g.][]{white_core_1978,jenkins_mass_2001,tinker_toward_2008} and the need for narrow ranges of halo masses for an optimal detection of the splashback radius. The final, merged binning is presented in Figure \ref{fig_binning}; Table \ref{tab_binning} present the cluster counts.

\subsubsection{Halo Masses}

Using clusters both detected in MaDCoWS2 and in Suyaev-Zel'dovich surveys \citep[Planck, South Pole Telescope and Atacama Cosmology Telescope surveys, see e.g.][]{planck_collaboration_planck_2014,bleem_galaxy_2015,hilton_atacama_2018} or in the Extended Roentgen Survey with an Imaging Telescope Array \citep[eROSITA,][]{bulbul_srgerosita_2024}, \citet{thongkham_massive_2024b} provide four scaling relations between $M_{500}$, $S\slash N$ and redshift. On average, these relations yield $M_{500}=2.6\pm 0.2 \times 10^{14}~\mathrm{M_\odot}$ and $M_{500}=4.6\pm 0.6 \times 10^{14}~\mathrm{M_\odot}$ for $S\slash N=5$ clusters at $z=0.57$ and $z=1.77$; for a $S\slash N=10$ cluster at $z=0.57$, the expected $M_{500}$ is $4.4\pm 0.4 \times 10^{14}~\mathrm{M_\odot}$. We stress however that masses obtained with these scaling relations are uncertain since there is a large scatter between $M_{500}$ and $S\slash N$. Furthermore, \citetalias{thongkham_massive_2026} inferred systematically lower halo masses using \citet{more_splashback_2015} relationship between $M_{200}$ and splashback radii. For more details on the halo masses of MaDCoWS2 clusters, we refers the reader to the discussion section of \citetalias{thongkham_massive_2026}.

\begin{deluxetable*}{cccccccc}
\tablecaption{Number of clusters in each redshift bin and $S\slash N$ subdivisions.}\label{tab_binning}
\tablehead{\multicolumn{2}{c}{Redshift} &\multicolumn{5}{c}{Cluster count} & \colhead{Total}\\
\colhead{Range} & \colhead{Mean} & \colhead{$5.0 \leq S\slash N < 5.25$} & \colhead{$5.25 \leq S\slash N < 5.91$} & \colhead{$5.91 \leq S\slash N < 7.16$} & \colhead{$7.16 \leq S\slash N < 9.23$} & \colhead{$S\slash N \geq 9.23$} & 
}
\startdata
$0.5\leq z < 0.65$ & $0.57$ & 4577 & 8620 & 7209 & 3795 & 964 & 25165\\
$0.65\leq z < 0.82$ & $0.72$ & 4094 & 7238 & 4893 & 2231 & merged & 18456\\
$0.82\leq z < 1.00 $ & $0.90$ & 3134 & 5362 & 3287 & 1288 & merged & 13071\\
$1.00\leq z < 1.20$ & $1.09$ & 3098 & 4766 & 2579 & 790 & merged & 11233\\
$1.20\leq z < 1.42$ & $1.30$ & 2429 & 3530 & 1941 & merged & merged & 7900\\
$1.42\leq z < 1.66$ & $1.53$ & 1609 & 2256 & 1043 & merged & merged & 4908\\
$1.66\leq z < 1.92$ & $1.77$ & 955 & 1657 & merged & merged & merged & 2612\\
\enddata
\tablecomments{``Merged'' means that a $S\slash N$ subdivision did not contain at least 750 clusters and was thus merged with a lower $S\slash N$ subdivision. ``Total'' refers to the sum of the different $S\slash N$ subdivisions within each redshift bin.}
\end{deluxetable*}

\subsection{WISE images}\label{ssec_wise_data}

WISE \citep[][]{wright_wide-field_2010} was a space-based survey observatory launched in 2010 and decommissioned in 2024. WISE had four infrared bands, which were centered at 3.4 $\mu$m (W1), 4.5 $\mu$m (W2), 12 $\mu$m (W3), and 22 $\mu$m (W4). A complete survey of the sky took 6 months, but the spacecraft emptied it reserve of cryogenic coolant after nine months of operations. Thus, there is only a single full coverage of the sky in W3 and W4. WISE continued its survey in W1 and W2 but was then placed in a hibernation state in February 2011. Reactivated in December 2013 under the name Near-Earth Object WISE \citep[NEOWISE;][]{mainzer_initial_2014}, WISE completed a total of 23 sky coverages in W1 and W2 before being decommissioned on August 8, 2024 due to orbital decay. WISE reentered the atmosphere on November 2, 2024.

In this work, we use the unblurred WISE coadded images (unWISE), made by \citet[][see also \citealt{lang_unwise_2014} and \citealt{meisner_full-depth_2017}]{meisner_deep_2017} with the W1 and W2 images taken prior to or during the seventh year of the NEOWISE mission. The unWISE coadds preserve the full resolution, contrary to the PSF-convolved images made by the WISE team \citep{masci_awaic_2009}. We downloaded each science and standard deviation cutout from the publicly available unWISE cutout service,\footnote{See \url{https://unwise.me/imgsearch/} and \url{https://doi.org/10.26131/irsa524}} and used the Legacy Survey viewer cutout service for mask cutouts. \footnote{\url{https://legacysurvey.org/viewer}}

\subsection{Image processing}\label{ssec_wise_pipeline}

The WISE cutouts are processed following a method very similar to the one described in \citetalias{trudeau_massive_2024}. The only significant change is the size of the image, here set to about $21^\prime.2 \times 21^\prime.2$ (equivalent to $8 \times 8$ Mpc at $z=0.5$). This larger size enables us to trace the total light stack profiles out to 3 Mpc from the cluster center, with a residual background measurement performed between 3 and 4 Mpc.

The unWISE images are organized in tiles; the cutout service returns as many science and standard deviation images as there are tiles overlapping with the requested cutout. In addition to the scalar background subtracted during the creation of the coadds \citep{lang_unwise_2014}, we perform an additional background subtraction on each individual (i.e. non-merged) cutout assuming the background behaves like a polynomial of the first order (i.e. a plane). The masking, which is done in three steps, is the same for each pair of W1 and W2 cutouts.

The unWISE masks include stellar diffraction spikes but insufficiently mask the stellar halos. We thus supplement the mask with a larger ellipse positioned on each point source of magnitude $W1\leq 15$. The second step consists of masking foreground extended galaxies more than 2.5 magnitudes brighter than the characteristic magnitudes (m*), assuming the luminosity function of \citet{mancone_formation_2010}. Finally, we mask photometric non-members, defining non-members as galaxies for which $| z_\mathrm{mode} - z_\mathrm{cluster} | > 3 \sigma_z$, where $z_\mathrm{mode}$ and $\sigma_z$ are the mode and standard deviation of the galaxy PDF. $z_\mathrm{cluster}$ is the best estimate of the cluster photometric redshift.

Should more than one cutout be necessary to reconstruct a $21^\prime.2 \times 21^\prime.2$ field of view, we use the {\tt reproject} python package with the {\tt reproject\_exact}\footnote{See the documentation at \url{https://reproject.readthedocs.io/en/stable/\#module-reproject}} algorithm to merge them. We perform this operation between the first and second step of the masking process. Standard deviation maps are added in quadrature.

Each pixel in our total light stacks is weighted by the inverse of its variance. To circularize the PSF, each pair of science and weight maps is rotated by a random multiple of 90\textdegree~before being stacked. We generate a thousand bootstrap simulations per bin, with replacement. These bootstraps are then used for splashback radii determination and error measurements.

\begin{figure*}
\begin{center}
\includegraphics[height=0.17\textheight]{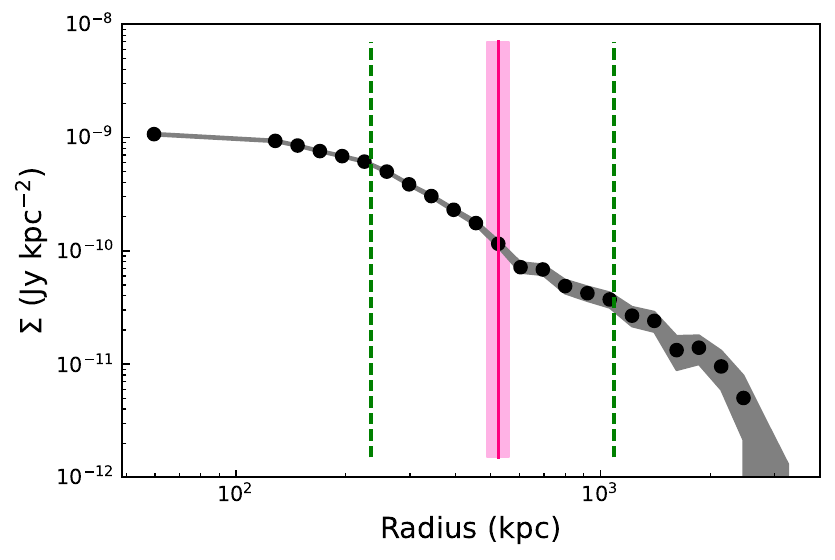}
\includegraphics[height=0.17\textheight]{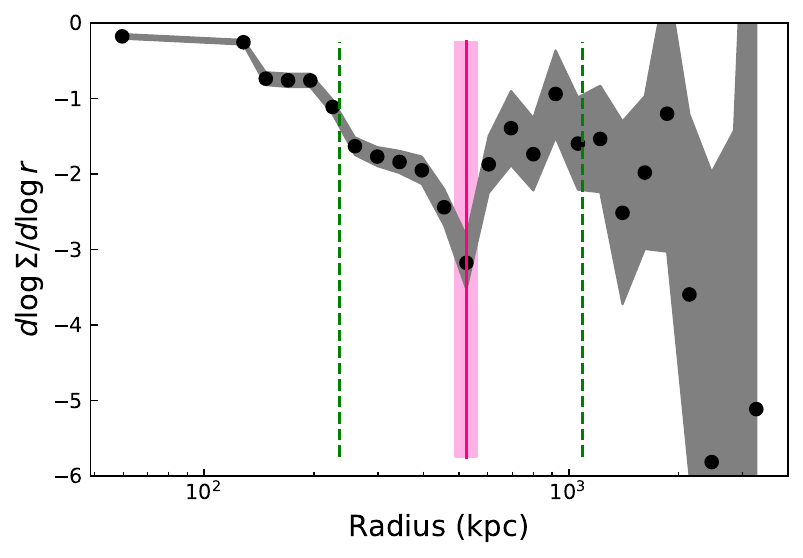}
\includegraphics[height=0.17\textheight]{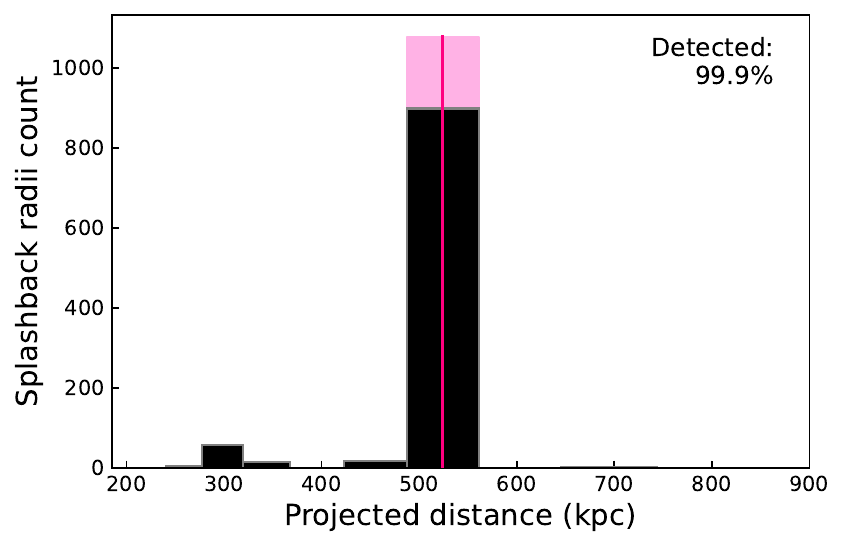}\\
\includegraphics[height=0.17\textheight]{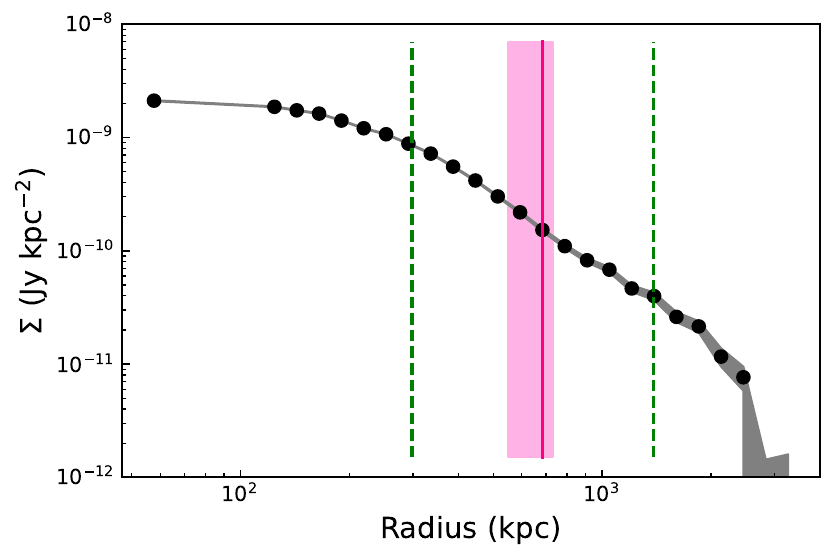}
\includegraphics[height=0.17\textheight]{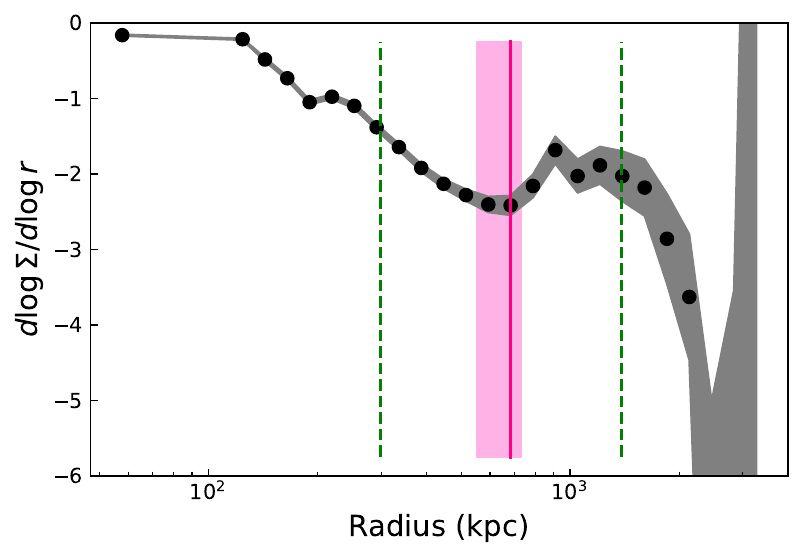}
\includegraphics[height=0.17\textheight]{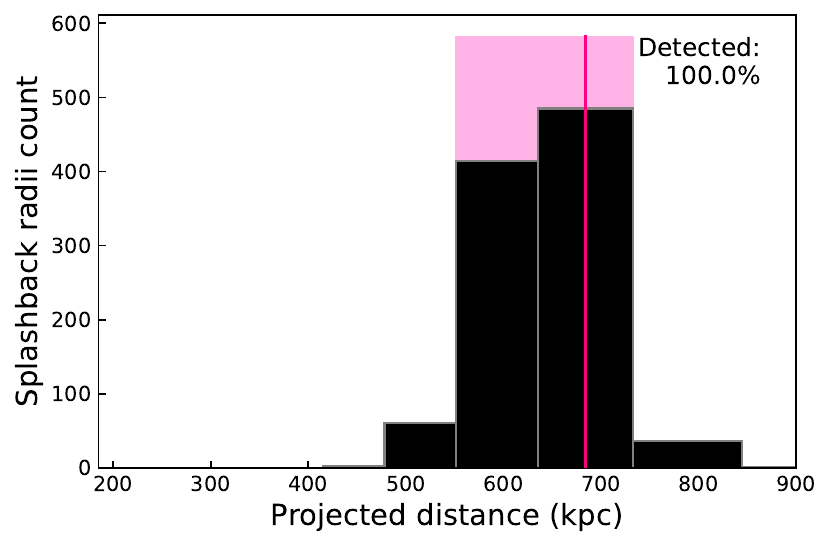}
\end{center}

\caption{Left: Non-cumulative mean surface brightness of the $\overline{z} = 1.77$, $5.0 \leq S\slash N < 5.25$ bootstraps (variance-weighted mean of W1 and W2). The confidence interval of the splashback radius is shaded in pink, and its median location indicated by a darker line. The edges of the prior are indicated in dark green. Center: Logarithmic slope of the surface brightness. Right: Histogram of the positions of the splashback radii in the W1+W2 bootstraps with the confidence interval shaded in pink. The text on the top right indicates the percentage of bootstraps with a detected feature. Bottom panels: same as the top panel, for the $\overline{z} = 1.09$, $5.25 \leq S\slash N < 5.91$ bootstraps.}
\label{fig_splashback_determination}
\end{figure*}

\begin{deluxetable*}{ccccccccc}
\tablecaption{Median splashback radii with errors, and percentage of bootstraps in which a caustic is detected.}\label{tab_splashback}
\tablehead{
\colhead{Mean redshift} & \multicolumn{2}{c}{$5.0 \leq S\slash N < 5.25$} & \multicolumn{2}{c}{$5.25 \leq S\slash N < 5.91$} & \multicolumn{2}{c}{$5.91 \leq S\slash N < 7.16$} & \multicolumn{2}{c}{$7.16 \leq S\slash N < 9.23$}\\
 & \colhead{(kpc)} & \colhead{(\%)} & \colhead{(kpc)} & \colhead{(\%)} & \colhead{(kpc)} & \colhead{(\%)} & \colhead{(kpc)} & \colhead{(\%)}}
\startdata
$0.57$ & $\ldots$ & $\ldots$ & $831^{+63}_{-171}$ & 99.5 & $\ldots$ & $\ldots$ & $\ldots$ & $\ldots$ \\
$0.72$ & $\ldots$ & $\ldots$ & $748^{+55}_{-231}$ & 94.5 & $867\pm 64$ & 93.9 & $\ldots$ & $\ldots$ \\
$0.90$ & $669\pm 48$ & 100.0 & $773^{+184}_{-152}$ & 99.9 & $773^{+56}_{-152}$ & 98.1 & $\ldots$ & $\ldots$ \\
$1.09$ & $\ldots$ & $\ldots$ & $684^{+49}_{-133}$ & 100.0 & $\ldots$ & $\ldots$ & $684\pm 49$ & 100.0 \\
$1.30$ & $601^{+42}_{-116}$ & 99.9 & $692^{+49}_{-133}$ & 100.0 & $521^{219}_{-101}$ & 99.0 & \multicolumn{2}{c}{merged}\\
$1.53$ & $\ldots$ & $\ldots$ & $456\pm 32$ & 100.0 & $525^{+101}_{-37}$ & 98.4 & \multicolumn{2}{c}{merged}\\
$1.77$ & $524\pm 37$ & 99.9 & $603^{+236}_{-42}$ & 100.0 & \multicolumn{2}{c}{merged} & \multicolumn{2}{c}{merged}\\
\enddata
\tablecomments{Measurements that do not follow the criteria listed in Section \ref{ssec_splashback} are omitted ($\ldots$). We do not show the $S\slash N \geq 9.23$ subdivision because it contains no reliable splashback detection.}
\end{deluxetable*}

\subsubsection{Residual background corrections}\label{sssec_back_sub}

As in \citetalias{trudeau_massive_2024}, the statistical contribution of interloping galaxies and imperfect background subtraction can result in a non-zero surface brightness at large radii. To correct for this residual background, we measure the surface brightness in annuli between 3 and 4 Mpc in the considered total light stack or bootstrap — and subtract this residual background from the surface brightnesses measured on that stack or bootstrap. We apply a similar correction to our flux measurements.

The dimensions of these background annuli were chosen as a compromise between the size of the stacks and the necessity of annuli far from the cluster center to avoid underestimating the cluster emission. We explore the impact of possible oversubtraction in Section \ref{sssec_back_sub_sys}.

\section{Results}\label{sec_results}

\subsection{Splashback radius measurements}\label{ssec_splashback}

Most authors define the splashback radius either as the point with the ``steepest slope'' or as a ``sharp break'' in the halo profile \citep[][see also \citealt{adhikari_splashback_2014,diemer_dependence_2014,deason_stellar_2021,gonzalez_discovery_2021}]{more_splashback_2015}. The breaks in our profiles are not very sharp or deep, but appear in both W1 and W2, and in most of our redshift and $S\slash N$ bins. We therefore use bootstrap simulations to assess the variability and reliability of our detections.

Our innermost aperture covers the five central pixels (within 45 to 60 kpc of the stack center, depending on redshift). Within this aperture the centering uncertainty results in a flat slope for the surface brightness profile. We then add 24 concentric annular apertures logarithmically spaced \citep[base ten logarithms, e.g.][]{gonzalez_discovery_2021}, out to a radius corresponding to 3.5 Mpc at the probed redshift.

Since W1 and W2 exhibit the same behavior, we maximise the S\slash N by combining the surface brightness measurements of each W1 bootstrap with a W2 bootstrap, using an inverse variance-weighted mean. The left panels of Figure \ref{fig_splashback_determination} show the median surface brightness profiles of the combined W1 and W2 bootstraps, for the $\overline{z} = 1.77$, $5.0 \leq S\slash N < 5.25$ and $\overline{z} = 1.09$, $5.25 \leq S\slash N < 5.91$ bins. The gray shaded regions correspond to the standard deviations of the combined profiles. The central panels show the median logarithmic slopes of the combined bootstraps.

In simulations \citep[e.g.][]{diemer_dependence_2014,diemer_dynamics-based_2023}, the splashback radius is usually located around $r_{200}$. We thus impose a prior corresponding to the $r_{200}$ of clusters with $M_{200}$ between $10^{13}$ and $10^{15}~M_\odot$. We select one tentative splashback radius per combined bootstrap in this prior, selecting the innermost ``dip'' when more than one steepening of the slope is present; the outermost variations are most likely driven by flux uncertainties. In some cases we do not detect any splashback radius in the combined bootstrap. These non-detections are taken into account to assess the reliability of our results.

The right panels of Figure \ref{fig_splashback_determination} show 
the distribution of inferred splashback radii within two redshift and $S\slash N$ subdivisions. Each bar correspond to an aperture and the height of the bar corresponds to the number of W1+W2 bootstrap where a caustic is detected in this aperture. The top right percentage indicate the fraction W1+W2 bootstrap simulations with a caustic detection. The shaded regions correspond to the 16th to 84th percentiles, taking into account the aperture sizes.

The $\overline{z} = 1.77$, $5.0 \leq S\slash N < 5.25$ and $\overline{z} = 1.09$, $5.25 \leq S\slash N < 5.91$ bins presented in Figure \ref{fig_splashback_determination} correspond to clear measurements: at least 99.9\% of the bootstraps have a dip and their distributions are well-behaved. However, in some bins the caustic distributions are less symmetric, or dips are detected in only a fraction of the bootstraps. To segregate the reliable detections from the more tentative ones we adopt the following rules:

\begin{enumerate}
    \item At least 90\% of the W1+W2 bootstrap derivatives must contain a ``dip;'' and
    \item Any secondary peaks in the radii distribution must be lower than a third of the primary peak height
\end{enumerate}

The caustic detections are presented in Table \ref{tab_splashback}. Measurements that do not respect all criteria listed above are not used in the analysis. We omit the highest $S\slash N$ bin ($S\slash N \geq 9.23$) in Table \ref{tab_splashback} because the total light stacks in that bin do not yield a reliable measurement of splashback radius. 
Since this bin is only defined for ($\overline{z} = 0.57$), we could have merged it with the $7.16\leq S\slash N < 9.23$ bin though we decided not to so to avoid having a very wide $S\slash N$ bin.

\subsection{Cluster profiles}\label{ssec_profiles}

Most splashback analyses based on stacked galaxy number counts infer the 3D location of the splashback radii from the best-fitting \citet{diemer_dependence_2014} model \citep[e.g.][]{baxter_halo_2017,zurcher_splashback_2019,xhakaj_how_2020,adhikari_probing_2021,rana_erosita_2023}. The \citet{diemer_dependence_2014} models combine an Einasto profile \citep[][see also \citealt{merritt_empirical_2006}]{einasto_construction_1965} describing the inner region and a power law to describe the infalling region. The transition region (i.e. the splashback radius) is modeled as another power law, with additional free parameters to control the truncation radius and how rapidly the slope varies.

In this work, we do not attempt to use \citet{diemer_dependence_2014} models to determine the 3D location of the splashback radii, because of the additional complexities of the miscentering and PSF convolution. For the analysis presented in Section \ref{sec_infalling}, we use Einasto profiles to compute aperture corrections — \citet{diemer_dependence_2014} models are too computationally expensive because of their many free parameters. The details of the aperture correction calculations are presented in Appendix \ref{sec_aperture_correction}.

\section{Discussion}\label{sec_discussion}
\subsection{Comparison with K. Thongkham et al.}\label{ssec_splash_compared}

\begin{figure}
\centering
\includegraphics[width=\columnwidth]{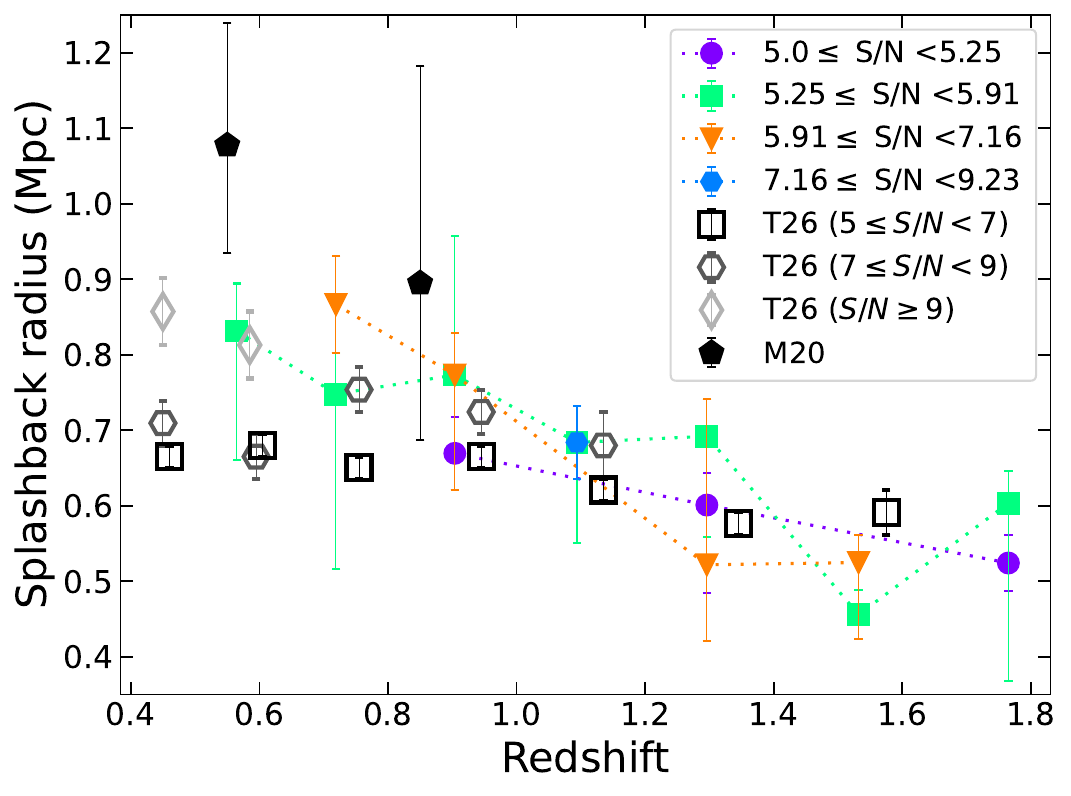}
\caption{Comparison between our splashback radii, color-coded by $S\slash N$ bins, and \citetalias{thongkham_massive_2026} 2D splashback radii (open symbols). For $z\geq 0.5$, we offset \citetalias{thongkham_massive_2026} values by $\Delta z=0.025$ to limit overlaps. We also include the two highest redshift bins of \citetalias{murata_splashback_2020} as filled black symbols.}
\label{fig_splashback_radii}
\end{figure}

\citetalias{thongkham_massive_2026} rely on a well-known method to detect splashback radii around MaDCoWS2 clusters \citep[e.g.][]{more_detection_2016,baxter_halo_2017,shin_mass_2021,rana_erosita_2023}. They compute the projected galaxy densities around the cluster centers using a Landy-Szalay estimator \citep{landy_bias_1993}. They then take the logarithmic derivative of the stacked galaxy density profiles to detect the splashback radius. \citetalias{thongkham_massive_2026} thus provides an independent determination of the splashback radii of MaDCoWS2 clusters, which is an invaluable benchmark with which to assess the performance of the detection technique presented here.

Figure \ref{fig_splashback_radii} presents a comparison between the projected splashback radius determined by \citetalias{thongkham_massive_2026} and our results. Despite our smaller $S\slash N$ subdivisions, our error bars overlap with \citetalias{thongkham_massive_2026} measurements. At low redshifts, \citetalias{thongkham_massive_2026} splashback radii exhibit a dependence on $S\slash N$, with splashback radii being larger for the highest $S\slash N$ bins. Two factors may explains why we do not observe this trend in our total light stacks. First, our measurements have larger error bars than \citetalias{thongkham_massive_2026} measurements. Second, our detection method performs better at high redshifts, a regime in which the range of possible cluster masses (and thus the number of $S\slash N$ bins) is restricted both by the detection limit of the MaDCoWS2 survey and by the increasing rarity of very massive halos at high redshift \citep[e.g.][]{tinker_toward_2008,watson_halo_2013,seppi_mass_2021}.

We also include the 2D splashback radii of \citetalias{murata_splashback_2020} (see Table 3) for the Hyper Suprime-Cam sample. While there are several splashback determinations in the literature, only a few cluster samples overlap with our redshift range \citep[][\citetalias{murata_splashback_2020}]{chang_splashback_2018,shin_measurement_2019,shin_mass_2021,giocoli_amico_2024}, and \citetalias{murata_splashback_2020} is one of the only works that provide the 2D values of their splashback radii. As shown by Figure \ref{fig_splashback_radii}, the \citetalias{murata_splashback_2020} $z\sim 0.55$ determination is significantly higher than our comparable values. However, their $z\sim 0.85$ splashback radius is consistent with our determinations and with \citetalias{thongkham_massive_2026} measurements. \citetalias{murata_splashback_2020} and \citetalias{thongkham_massive_2026} use similar methods, which suggests that the discrepancy at $z\sim 0.55$ is driven by the difference in the cluster sample rather than systematic errors. Specifically, the larger splashback radius of \citetalias{murata_splashback_2020} suggests that their richness cut of 15 might correspond to a larger mass than $S\slash N=5$ in MaDCoWS2.

The consistency of our measurements with those of \citetalias{thongkham_massive_2026} suggests that our results are robust to effects pertaining to the galaxy populations. \citetalias{thongkham_massive_2026} galaxy counts rely on galaxies bright enough to be individually detected in CatWISE2020 and DECaLS catalogs. In contrast, our variance-weighted total light stacks are sensitive to the flux of all clusters members. However, because we averaging over the flux, the dominant contribution probably comes from the galaxies close to $L^\ast$, the Schechter function characteristic luminosity \citep{schechter_analytic_1976}.

The effects of these different galaxy populations on the location of the splashback radii are unclear. 
\citet[][see also \citealt{dacunha_connecting_2022}]{oneil_impact_2022} noted that less massive galaxies are more accurate tracers of the underlying dark matter distribution than more massive galaxies, which can underpredict the splashback radius by as much as 25\%. This suggests that both \citetalias{thongkham_massive_2026} and our total light stack might be biased toward lower values. However, the relative impact of this bias at a given redshift depends on how CatWISE2020 and DECaLS magnitude limits compare with $L^\ast$ in the MaDCoWS2 sample.

In contrast, several studies have shown that quiescent galaxies \citep[][\citetalias{murata_splashback_2020}]{adhikari_probing_2021,dacunha_connecting_2022,oneil_impact_2022} are better tracers of the splashback radius than star-forming galaxies. For equivalent masses, star-forming galaxies are more luminous than quiescent galaxies and would thus possibly play a bigger role in our estimates than in \citetalias{thongkham_massive_2026}. We would however expect these galaxies to bias our estimates toward lower values than the measurements of \citetalias{thongkham_massive_2026} (see e.g. \citealt[][]{dacunha_connecting_2022,oneil_impact_2022}), an effect we do not observe. 

\subsubsection{ICL contribution}\label{sssec_ICL_adv}

The analysis in \citetalias{thongkham_massive_2026} and our analysis also differ with regards to the potential impact of the ICL. Total light stacks like ours are sensitive to the total light of the clusters; the contribution of the ICL to the total light budget is however uncertain. Estimates of the ICL contribution to the stellar mass or luminosity vary, spanning 10 to 35\% \citep{contini_formation_2014,ragusa_does_2023,montenegro-taborda_stellar_2025}. There is a possible inverse relationship with halo mass, with larger ICL contributions more common in low-mass clusters \citep[e.g.][]{gonzalez_census_2007,ragusa_does_2023,contreras-santos_characterising_2024,montenegro-taborda_stellar_2025}.

The impact of the ICL component on the splashback radius location is however expected to be minimal. Using simulations, \citet{deason_stellar_2021} find that the ICL and dark matter splashback radius locations are in good agreement, albeit with the caveat that low accretion rates can result in double caustics in the ICL profile.

\subsection{Systematic errors}\label{ssec_systematics}

The general agreement between \citetalias{thongkham_massive_2026} and our method also implies that most sources of systematic errors have either a similar impact on both methods or are subdominant to the statistical uncertainties. Nevertheless, for completeness, we provide in this section a discussion of two of the main sources of systematics associated with the total light stacking method: size evolution (both apparent and intrinsic) and background estimations. We also briefly discuss other potential systematics.

\subsubsection{Intrinsic and apparent cluster size evolution}\label{sssec_ang_size}

In Section \ref{ssec_madcows2}, we decided to use a non-linear binning scheme as a compromise between cluster counts and cluster size homogeneity. As a sanity check, we tested wider $S\slash N$ bins, which resulted in shallower splashback radii detections. 
While this result, as well as the results of \citetalias{thongkham_massive_2026}, indicate a possible relationship between cluster mass and halo extent, the literature is inconclusive on this topic. The \citet{alberts_clusters_2022} review article compiled concentration measurements ($c=r_{200}/r_s$, where $r_s$ is the scale factor of a Navarro-Frenk-White model) from different sources and found no evidence of evolution with mass or redshift. Without a clear relationship between halo masses and characteristic scales, we cannot predict and correct the effect of intrinsic size variations within a bin.

We can however estimate the impact of apparent size variations within a bin. To do so, we generate a set of total light stacks with ``rescaled'' pixels. We multiply the pixel size in each individual cutout header by a correction factor, corresponding to the angular size of a kiloparcsec at the redshift of the cluster, divided by the angular size of a kiloparsec at the mean redshift of the corresponding bin. We then resample the pixels of each image back to a common value using the package {\tt reproject}. We generate bootstrap simulations for these images and measure their splashback radius distributions as described in Section \ref{ssec_splashback}.

We find that the original and rescaled splashback distributions are not identical, but have largely overlapping confidence intervals. There is no indication of one set performing significantly better than the other in terms of detection rates or confidence intervals, nor is there a systematic shift toward smaller or larger median splashback radii. These observations suggest that while there might be some apparent size variations within a redshift bin, the imprecision of the photometric redshifts limits the usefulness of a rescaling correction. In addition, rescaled total light stacks would be difficult to compare to a profile model (see Appendix \ref{sec_aperture_correction}) because of the PSF varying from image to image within the stack.

\subsubsection{Background subtraction}\label{sssec_back_sub_sys}

\begin{figure*}
\begin{center} 
\includegraphics[height=0.17\textheight]{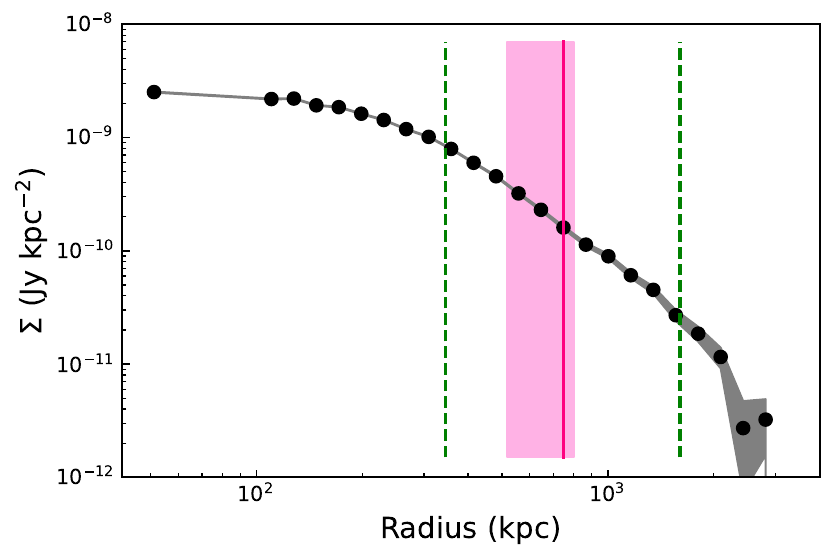}
\includegraphics[height=0.17\textheight]{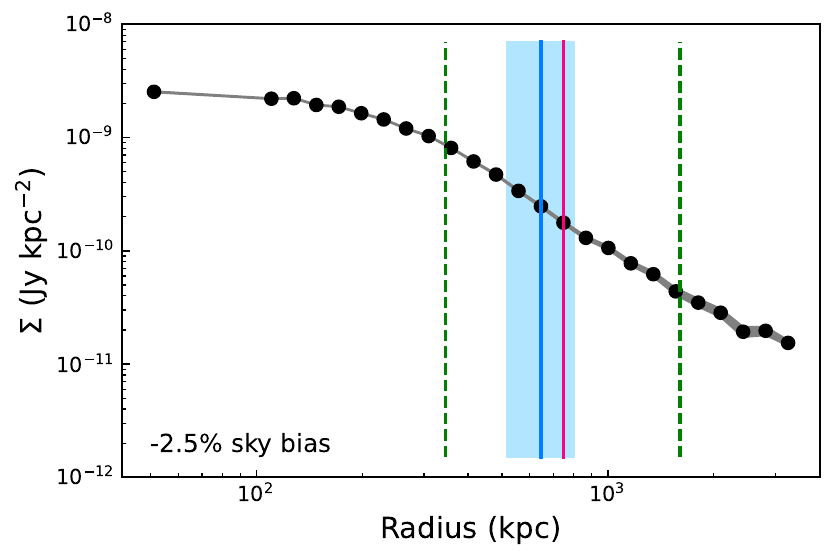}
\includegraphics[height=0.17\textheight]{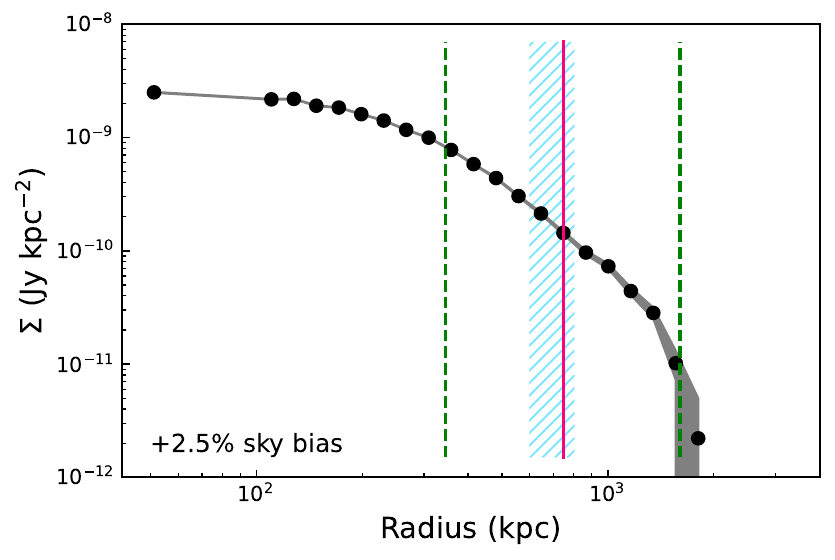}\\
\includegraphics[height=0.17\textheight]{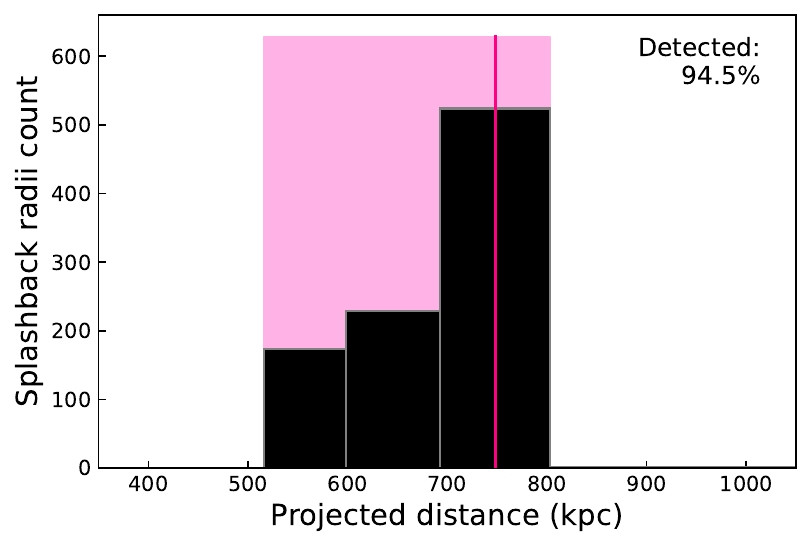}
\includegraphics[height=0.17\textheight]{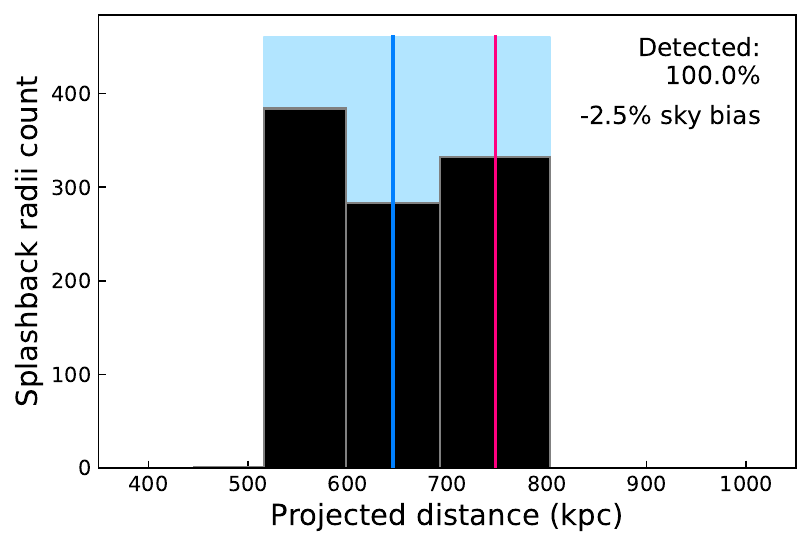}
\includegraphics[height=0.17\textheight]{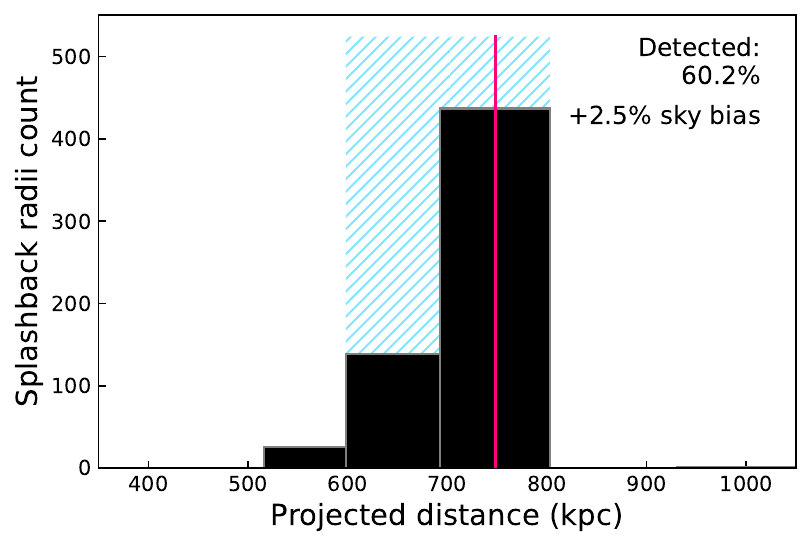}
\end{center}

\caption{Splashback radius measurements for the $\overline{z} = 0.72$, $5.25 \leq S\slash N < 5.91$ bootstraps. The confidence interval of the splashback radius is shaded in pink if the determination is reliable (see Section \ref{ssec_splashback}) and hatched if it is not. Left: Nominal background determination. Center: Same as left panel, but introducing a systematic 2.5\% undersubtraction of the residual background. The new confidence interval is shaded or hatched in blue, with the new median indicated by a blue line. The median of the original distribution is indicated by a pink line. Right: mean surface brightness with a systematic 2.5\% oversubtraction of the residual background. The new confidence interval (in blue) is not deemed reliable though the new median and original median have the same value, i.e. the blue line is under the pink line.}
\label{fig_splashback_backgrounds}
\end{figure*}

The residual background correction discussed in Section \ref{sssec_back_sub} corrects for systematic over or undersubtraction of the background during processing of the individual images. An accurate background correction is critical for an accurate detection of the splashback radius.

In this paper, we choose to measure the background residual correction in 3 to 4 Mpc annuli around the total light stack centroid rather than making separate sky surface brightness stacks. This choice is motivated by computing efficiency and convenience: generating slightly larger total light stacks is easier than generating, masking, and stacking as many sky images as there are stacks.

This annular strategy could however result in systematic oversubtraction of the residual background if the infalling region extends beyond 3 Mpc. We explore the impact of this potential systematic by multiplying the surface brightness measured in our background annuli by 0.975 or 1.025 — i.e. by introducing sky biases of 2.5\%. We also test deviations of 5\%. Figure \ref{fig_splashback_backgrounds} shows a comparison between the original profiles and splashback radii distributions of the $\overline{z} = 0.72$, $5.25 \leq S\slash N < 5.91$ bin (left panels) and the versions with background residual underestimation (center) and overestimation (right).

In the underestimated subtraction panels, the transition between the cluster and remaining background results in a change of the slope, very similar to what occurs at
the splashback radius. Thus, if that transition occurs close to the expected location of the splashback radius it might be mistaken for it. This is what happens in the central panels of Figure \ref{fig_splashback_backgrounds}: the splashback radius distribution is biased toward lower values because this is where the remaining background starts to dominate in the majority of the bootstraps realizations.

In the right panels of Figure \ref{fig_splashback_backgrounds}, the residual background is overestimated, resulting in a negative surface brightness at large radii and a logarithmic derivative falling steeply toward infinity. The main difference here is that this steep fall does not mimic the behavior of the splashback radii. Thus, a systematic oversubtraction will limit splashback radius counts — the splashback radius will not be detected if it lies in a region with negative surface brightness — but will have a limited impact on the shape of the splashback distribution. In the right panels of Figure \ref{fig_splashback_backgrounds} a splashback radius is detected in only 60\% of the bootstrap realizations. The resulting distribution thus fails to reach the minimal criteria for a reliable detection (see Section \ref{ssec_splashback}), but the shape remains qualitatively similar to the original distribution.

\subsubsection{Other systematics}\label{sssec_other_systematics}

Another possible systematic is miscentering, i.e., the difference between the cluster center as detected by the MaDCoWS2 survey and the real center of the dark matter distribution. This effect impacts both \citetalias{thongkham_massive_2026} and our method. \citet{thongkham_massive_2024b} measured the positional offset between MaDCoWS2 and six other surveys. They found that the standard deviation of the offset between MaDCoWS2 and most comparable surveys is 0.2 Mpc. \citetalias{thongkham_massive_2026} investigated this uncertainty and found that it is one of the dominant contributions to their uncertainty budget. While we do not perform a similar decomposition, we expect our bootstrap-based splashback distribution to reflect the scatter caused by miscentering — and we note that the uncertainties computed by \citetalias{thongkham_massive_2026} are of the order of 15 to 45 kpc, significantly smaller than the apertures we use to sample surface brightness.

\subsection{Superior sensitivity of stacking at high redshift}\label{ssec_high_z_adv}

The vast majority of splashback radius detections in the literature are at $z\lesssim 0.7$. Splashback radii are usually detected via weak lensing \citep[e.g.][]{chang_splashback_2018,contigiani_weak_2019,fong_first_2022,giocoli_amico_2024,shin_mass_2021} or galaxy counts measurements \citep[e.g.][]{adhikari_observing_2016,adhikari_probing_2021,baxter_halo_2017,chang_splashback_2018,more_detection_2016,rana_erosita_2023,shin_mass_2021,xu_measurement_2024}. Weak lensing estimates require deep images at high resolution to accurately measure the distortion of background galaxies. While Euclid and JWST images are now available for high-redshift clusters, shear lensing systematics are still a limitation since they remain poorly understood in the near infrared \citep[e.g.][]{finner_constraining_2020}. For galaxy counts studies, the depth of the existing galaxy catalogs are the main limiting factors.

\citetalias{thongkham_massive_2026} are the first to apply galaxy counts to high redshift galaxy clusters, but even they struggle to recover splashback radii at $z > 1.5$ because of the limited number of high-redshift galaxies bright enough to be detected in their DECaLS+CatWISE2020 galaxy catalog. In contrast, our splashback radius detections tend to have smaller error bars at high redshift, despite our larger and less populated bins.

We suggest that the sensitivity of total light stacks to the full light budget of clusters may represent an advantage at high redshift. Unlike other techniques, total light stacks do not rely on multiple galaxy detections in individual observations. Our signal-to-noise is instead approximately proportional to the square root of the number of stacked images \citep{kelly_60_1990,garn_radio_2009,bourne_evolution_2011}. Thus, provided that we have a sufficient number of clusters, we can detect a signal even without a significant number of individually detected galaxies in the input images.

However, our good performance at high redshift might be influenced by the depth and location of the caustics. Halo accretion rates are expected to be higher at high redshift than at later epochs \citep[e.g.][]{fakhouri_nearly_2008,fakhouri_merger_2010,wang_assembly_2011,chiang_ancient_2013,diemer_splashback_2017-1,dong_universal_2022}. High accretion rates tend to produce deep caustics closer to the cluster center \citep[e.g.][]{diemer_dependence_2014,deason_stellar_2021} — two conditions that favor detection with total light stacks. 
This sensitivity to sources too faint to be individually detected also make total light stacks a useful tool for galaxy evolution studies. In the following Section, we explore how splashback radii can be used to as a probe of galaxy environment.

\section{Application: the stellar mass in the infalling region}\label{sec_infalling}

\begin{figure}
\centering
\includegraphics[width=\columnwidth]{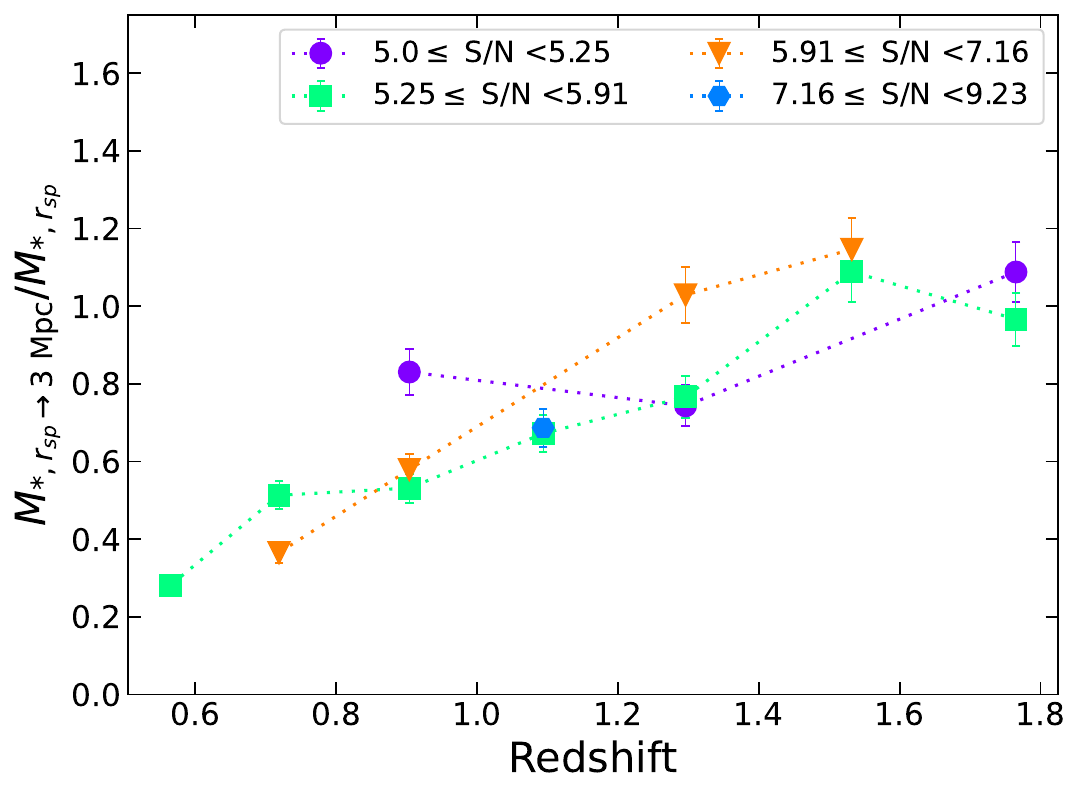}
\caption{The stellar mass (not deprojected) beyond the splashback radius (out to 3 Mpc) divided by the stellar mass within the splashback radius. The stellar mass in the outskirts increases with redshift with respect to the central mass.}
\label{fig_splashback_aperture}
\end{figure}

While there are extensive studies on galaxy population in clusters and their outskirts, the definitions of ``outskirts'' or ``infall region'' in the literature usually relies on $r_{200}$ \citep[e.g.][]{baxter_gogreen_2022,werner_satellite_2022,brambila_examining_2023,lopes_role_2024} rather than on a physical radius like the splashback radius. In this section, we present how our splashback determination technique could be used to
remediate to that problem. Using the W1 and W2 bootstrap realizations generated for the splashback radius measurements, we determine the (projected) fraction of the stellar mass within the infall region and its evolution with redshift.

Figure \ref{fig_splashback_aperture} shows the stellar mass in a cylindrical aperture between splashback radius and a projected radius of 3 Mpc, divided by the stellar mass within the projected splashback radius. We find that the fraction of stellar mass within the outermost aperture increases with redshift. There is no significant difference between the $S\slash N$ subdivisions, which suggests that the increase is driven by larger stellar masses in the infall region at high redshift, not by the halo mass function of the MaDCoWS2 catalog.

To create Figure \ref{fig_splashback_aperture}, we measured the W1 and W2 flux densities in two apertures: within the detected splashback radius and between the splashback radius and 3 Mpc. The flux densities in fixed apertures are presented in Appendix \ref{sec_masses}. In each bootstraps pair, the flux was measured using the location of the projected splashback radius in this particular pair (see Section \ref{ssec_splashback}); pairs without any splashback detection were omitted. We then applied a random aperture corrections (see Appendix \ref{sec_aperture_correction}) to each flux measurement. The random aperture correction is drawn from a normal distribution centered on the best aperture correction estimate, with its standard deviation corresponding to the largest error bar. 

To get the aperture-corrected flux beyond the splashback radius, we then subtracted the aperture-corrected inner flux from a 3 Mpc wide circular aperture. Fluxes are transformed into stellar masses using {\tt CIGALE} \citep{burgarella_star_2005,noll_analysis_2009,boquien_cigale_2019} assuming a delayed $\mathrm{\tau}$-model with a characteristic time of 1 Myr and a formation redshift of 3.

The above calculation could be improved by taking into account projection effects. We are also lacking flux measurements in multiple bands, which could be used to put constraints on the star formation history. These improvements are beyond the scope of this paper, but might be possible with the new generation of space-based survey telescopes such as Euclid \citep{euclid_collaboration_euclid_2025} and the \textit{Nancy Grace Roman Space Telescope} \citep{akeson_wide_2019}.

\section{Conclusions}\label{sec_conclusion}

We report the first detection of splashback radii in total light stacks. Our stacks are assembled from a sample of 83,345 galaxy clusters drawn from the second data release of MaDCoWS2 \citep{thongkham_massive_2024b}, and follow a very similar analysis method to \citetalias{trudeau_massive_2024}. We measure splashback radii in bootstrap realizations. We then take the median and the region between the 16th and 84th percentiles as our best value and confidence interval respectively. We compare our results to \citetalias{thongkham_massive_2026} measurements, which are based on the same cluster sample but measured using cluster-galaxy cross-correlations. We explore the main systematics and draw the following conclusions:
\begin{itemize}
    \item Our splashback radii are in good agreement with \citetalias{thongkham_massive_2026} measurements, indicating that our detections are genuine. The agreement between both results also suggest that total light stacks' sensitivity to different galaxy populations, compared to cross-correlation methods, has only a limited impact on our measurements.
    \item Our attempt to rescale the clusters according to their photometric redshifts did not improve our results, implying that the change of angular scale within a total light stack has a limited impact on our results. 
    \item Our measurements are very sensitive to background estimation. Care must be taken to avoid underestimating the background lest it bias measurements toward smaller radii. Background overestimation reduces the significance of the detection but does not change its value.
    \item Total light stacks are a useful tool for investigating splashback radii at high redshifts. Unlike galaxy count techniques, total light stacks are sensitive to contributions by galaxies too faint to be detected in individual images. Total light stacks also might be useful for analyses that combine galaxy evolution and cluster structure, such as an investigation of the galaxy population in cluster outskirts.
    \item We present a potential application of our splashback determination method to galaxy population studies. Using our calculated 2-dimensional splashback radius as the boundary between the clusters and their infall regions, we show that the fraction of stellar mass within the infall region appear to increases with redshift. This result remains however a proof of concept; more robust results will need multiwavelength photometry to improve stellar modeling, and will need to account for projection effects.
    
\end{itemize}


\begin{acknowledgments}

We thank Dustin Lang for making WISE mask cutouts available on the Legacy viewer website. This material is based upon work supported by the National Science Foundation under Grant No. 2108367. The authors acknowledge University of Florida Research Computing for providing computational resources and support that have contributed to the research results reported in this publication. The work of P.R.M.E. and D.S. was carried out at the Jet Propulsion Laboratory, California Institute of Technology, under a contract with the National Aeronautics and Space Administration (80NM0018D0004). The unWISE coadded images and catalog are based on data products from the \textit{Wide-field Infrared Survey Explorer}, which is a joint project of the University of California, Los Angeles, and the Jet Propulsion Laboratory/California Institute of Technology, and \textit{NEOWISE}, which is a project of the Jet Propulsion Laboratory/California Institute of Technology. WISE and \textit{NEOWISE} are funded by the National Aeronautics and Space Administration.

\end{acknowledgments}

\begin{contribution}



AT ran the analysis, wrote and submitted the manuscript. AHG provided supervision, obtained the funding and suggested significant improvements to the structure and the scope of the article. KT provided an independent proof of the concept, built the MaDCoWS2 catalog and provided the photometric PDF used by AT's codes.
Every other authors edited the manuscript and suggested improvements. Every author contributed to the MaDCoWS2 catalog.

\end{contribution}

\vspace{5mm}
\facilities{WISE}

\software{Astropy \citep[\hspace{-6pt}][]{astropy_collaboration_astropy_2013,astropy_collaboration_astropy_2018,astropy_collaboration_astropy_2022},
Cigale \citep{boquien_cigale_2019},
Colossus \citep{diemer_colossus_2018,diemer_accurate_2019},
Matplotlib \citep{hunter_matplotlib_2007}, 
NumPy \citep{harris_array_2020},
Reproject \citep{robitaille_reproject_2020,robitaille_astropyreproject_2023},
SciPy \citep{virtanen_scipy_2020},
Sep \citep{barbary_sep_2018}
}

\appendix

\section{Aperture corrections}\label{sec_aperture_correction}

To compute aperture corrections (i.e. correction factors accounting for PSF blurring and the miscentering of the clusters), we choose to use Einasto profiles as the best compromise between accuracy and computational resources. Einasto profiles more accurately describe the structure of dark matter halos than Navarro-Frenk-White profiles \citep[e.g.][]{gao_redshift_2008,navarro_diversity_2010,ludlow_density_2011}. However, they do not integrate the splashback radius (nor do the Navarro-Frenk-White profiles).

We compute the aperture corrections in the same way as in \citetalias{trudeau_massive_2024}. We generate two sets of models: one representing a perfect total light stack and another with miscentering and PSF blurring. The blurred models are created by stacking 100 iterations of Einasto profiles (2D projection) convolved with the PSF. The center of each iteration center is determined by a random distance from the stack center drawn from a 2D normal distribution and an angle drawn from an uniform distribution. The blurred stacks are then compared to the measured profiles using a $\chi^2$ minimization. Besides the change of the models, we also updated the miscentering standard deviation to better reflect the typical centering uncertainty computed by \citet{thongkham_massive_2024b}.

\section{Fluxes and stellar masses in fixed apertures} \label{sec_masses}

\begin{deluxetable}{cccccccccc}
\tablecaption{Aperture-corrected W1 and W2 flux densities and 2$\sigma$ limits, organized as a function of redshifts, $S\slash N$, apertures and bands.}\label{tab_fluxes}
\tablehead{\colhead{Mean redshift} & \colhead{$S\slash N$} & \multicolumn{2}{c}{$f_{1~\mathrm{Mpc}}$} & \multicolumn{2}{c}{$f_{1~\mathrm{Mpc}\rightarrow 3~\mathrm{Mpc}}$} & \multicolumn{2}{c}{$f_{r_{sp}}$} & \multicolumn{2}{c}{$f_{r_{sp}\rightarrow 3~\mathrm{Mpc}}$}\\
& & \colhead{W1} & \colhead{W2} & \colhead{W1} & \colhead{W2} & \colhead{W1} & \colhead{W2} & \colhead{W1} & \colhead{W2}\\
& & \colhead{(mJy)} & \colhead{(mJy)} & \colhead{(mJy)} & \colhead{(mJy)}& \colhead{(mJy)} & \colhead{(mJy)}& \colhead{(mJy)} & \colhead{(mJy)}}
\startdata
$0.57$ & $5.0 \leq S\slash N < 5.25$ & $1.387\pm 0.022$ & $0.920\pm 0.022$ & $<0.17$ & $<0.17$ & $\ldots$ & $\ldots$ & $\ldots$ & $\ldots$ \\
& $5.25 \leq S\slash N < 5.91$ & $1.594\pm 0.016$ & $1.066\pm 0.016$ & $0.289_{-0.060}^{+0.061}$ & $0.200\pm 0.061$ & $1.491\pm 0.059$ & $0.979\pm 0.046$ & $0.406\pm 0.097$ & $0.291\pm 0.086$\\
& $5.91 \leq S\slash N < 7.16$ & $2.030\pm 0.018$ & $1.347\pm 0.018$ & $0.396\pm 0.070$ & $0.218\pm 0.070$ & $\ldots$ & $\ldots$ & $\ldots$ & $\ldots$ \\
& $7.16 \leq S\slash N < 9.23$ & $2.770\pm 0.028$ & $1.839\pm 0.028$ & $1.045_{-0.100}^{+0.099}$ & $0.737\pm 0.099$ & $\ldots$ & $\ldots$ & $\ldots$ & $\ldots$ \\
& $S\slash N \geq 9.23$ & $4.197\pm 0.053$ & $2.807\pm 0.053$ & $1.68\pm 0.18$ & $1.18\pm 0.18$ & $\ldots$ & $\ldots$ & $\ldots$ & $\ldots$ \\
\hline
$0.72$ & $5.0 \leq S\slash N < 5.25$ & $1.271\pm 0.019$ & $0.823\pm 0.019$ & $0.291_{-0.080}^{+0.072}$ & $0.203_{-0.076}^{+0.073}$ & $\ldots$ & $\ldots$ & $\ldots$ & $\ldots$ \\
& $5.25 \leq S\slash N < 5.91$ & $1.521\pm 0.015$ & $0.995\pm 0.015$ & $0.434\pm 0.053$ & $0.327\pm 0.053$ & $1.346\pm 0.090$ & $0.855\pm 0.066$ & $0.64\pm 0.11$ & $0.492\pm 0.094$\\
& $5.91 \leq S\slash N < 7.16$ & $1.879\pm 0.018$ & $1.206\pm 0.018$ & $0.557\pm 0.065$ & $0.339\pm 0.065$ & $1.781\pm 0.082$ & $1.133\pm 0.060$ & $0.66\pm 0.11$ & $0.407\pm 0.090$\\
& $S\slash N \geq 7.16$ & $2.875\pm 0.029$ & $1.845\pm 0.029$ & $0.88\pm 0.10$ & $0.55\pm 0.10$ & $\ldots$ & $\ldots$ & $\ldots$ & $\ldots$ \\
\hline
$0.90$ & $5.0 \leq S\slash N < 5.25$ & $1.103\pm 0.020$ & $0.827\pm 0.020$ & $0.527_{-0.071}^{+0.070}$ & $0.504_{-0.071}^{+0.070}$ & $0.940\pm 0.043$ & $0.678\pm 0.036$ & $0.707\pm 0.089$ & $0.669\pm 0.089$\\
& $5.25 \leq S\slash N < 5.91$ & $1.253\pm 0.014$ & $0.925\pm 0.014$ & $0.449_{-0.055}^{+0.054}$ & $0.413_{-0.055}^{+0.054}$ & $1.146\pm 0.065$ & $0.829\pm 0.059$ & $0.553\pm 0.093$ & $0.506\pm 0.085$\\
& $5.91 \leq S\slash N < 7.16$ & $1.592\pm 0.019$ & $1.162\pm 0.019$ & $0.603_{-0.071}^{+0.073}$ & $0.487_{-0.071}^{+0.072}$ & $1.433\pm 0.087$ & $1.024\pm 0.073$ & $0.78\pm 0.12$ & $0.65\pm 0.11$\\
& S\slash N $\geq$ 7.16 & $2.384\pm 0.033$ & $1.739\pm 0.033$ & $0.74\pm 0.12$ & $0.64\pm 0.12$ & $\ldots$ & $\ldots$ & $\ldots$ & $\ldots$ \\
\hline
$1.09$ & $5.0 \leq S\slash N < 5.25$ & $0.903\pm 0.018$ & $0.796\pm 0.018$ & $0.274_{-0.057}^{+0.060}$ & $0.324_{-0.059}^{+0.061}$ & $\ldots$ & $\ldots$ & $\ldots$ & $\ldots$ \\
& $5.25 \leq S\slash N < 5.91$ & $1.056_{-0.014}^{+0.013}$ & $0.923_{-0.014}^{+0.013}$ & $0.391\pm 0.049$ & $0.355\pm 0.049$ & $0.900\pm 0.041$ & $ 0.760\pm 0.039$ & $0.577\pm 0.071$ & $0.541\pm 0.070$\\
& $5.91 \leq S\slash N < 7.16$ & $1.348\pm 0.020$ & $1.179\pm 0.020$ & $0.522_{-0.066}^{+0.071}$ & $0.525_{-0.068}^{+0.070}$& $\ldots$ & $\ldots$ & $\ldots$ & $\ldots$ \\
& S\slash N $\geq$ 7.16 & $1.887\pm 0.038$ & $1.632\pm 0.038$ & $0.69_{-0.14}^{+0.13}$ & $0.77_{-0.14}^{+0.13}$ & $1.62\pm 0.10$ & $1.349\pm 0.095$ & $0.98\pm 0.18$ & $1.06\pm 0.18$\\
\hline
$1.30$ & $5.0 \leq S\slash N < 5.25$ & $0.708\pm 0.017$ & $0.771\pm 0.017$ & $0.260_{-0.067}^{+0.066}$ & $0.300_{-0.067}^{+0.066}$ & $0.579\pm 0.021$ & $0.607\pm 0.24$ & $0.405\pm 0.075$ & $0.479\pm 0.081$\\
& $5.25 \leq S\slash N < 5.91$ & $0.806\pm 0.015$ & $0.859\pm 0.015$ & $0.344\pm 0.052$ & $0.420_{-0.052}^{+0.053}$ & $0.683\pm 0.035$ & $0.703\pm 0.039$ & $0.476\pm 0.066$ & $0.589\pm 0.069$\\
& $S\slash N \geq 5.91$ & $1.092\pm 0.020$ & $1.149\pm 0.020$ & $0.477_{-0.081}^{+0.076}$ & $0.585_{-0.081}^{+0.076}$ & $0.810\pm 0.088$ & $0.811\pm 0.099$ & $0.76\pm 0.12$ & $0.92\pm 0.14$\\
\hline
$1.53$ & $5.0 \leq S\slash N < 5.25$ & $0.524\pm 0.020$ & $0.656\pm 0.020$ & $0.228_{-0.075}^{+0.072}$ & $0.323_{-0.076}^{+0.073}$ & $\ldots$ & $\ldots$ & $\ldots$ & $\ldots$ \\
& $5.25 \leq S\slash N < 5.91$ & $0.566\pm 0.017$ & $0.712\pm 0.017$ & $0.207_{-0.069}^{+0.068}$ & $0.337_{-0.071}^{+0.069}$ & $0.393\pm 0.026$ & $0.472\pm 0.035$ & $0.377\pm 0.073$ & $0.575\pm 0.087$\\
& $S\slash N \geq 5.91$ & $0.812\pm 0.025$ & $0.966\pm 0.025$ & $0.362_{-0.089}^{+0.090}$ & $0.515\pm 0.091$ & $0.577\pm 0.078$ & $0.675\pm 0.093$ & $0.62\pm 0.12$ & $0.82\pm 0.14$\\
\hline
$1.77$ & $5.0 \leq S\slash N < 5.25$ & $0.389\pm 0.027$ & $0.536\pm 0.027$ & $<0.25$ & $0.254_{-0.116}^{+0.096}$ & $0.296\pm 0.024$ & $0.391\pm 0.033$ & $0.346\pm 0.094$ & $0.40\pm 0.11$\\
& S\slash N $\geq$ 5.25 & $0.524\pm 0.021$ & $0.680\pm 0.021$ & $0.239_{-0.072}^{+0.071}$ & $0.334\pm 0.072$ & $0.404\pm 0.055$ & $0.521\pm 0.070$ & $0.380\pm 0.091$ & $0.52\pm 0.11$\\
\enddata
\tablecomments{If the splashback radii is not reliable, columns $f_{r_{sp}}$ and $f_{r_{sp}\rightarrow 3~\mathrm{Mpc}}$ are left empty.}
\end{deluxetable}

\begin{figure}
\begin{center}
\includegraphics[width=\textwidth]{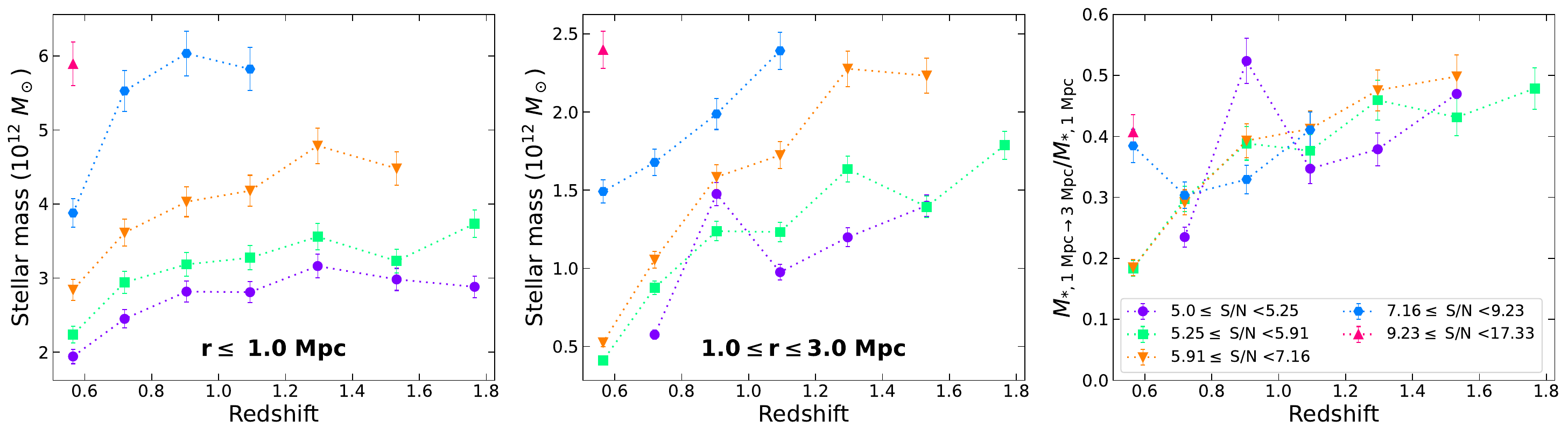}
\end{center}

\caption{Left: Stellar masses in a circular aperture of 1 Mpc radius, as a function of redshift, color-coded by $S\slash N$ bin. The stellar mass is computed assuming a delayed $\mathrm{\tau}$-model with a 1 Myr characteristic time and a formation redshift of $z_\mathrm{form}=3$. Middle: Stellar masses in an annular aperture between 1 and 3 Mpc, assuming the same star formation history. Right: Stellar mass between 1 and 3 Mpc divided by the stellar mass within 1 Mpc of the stack center. The fraction of stellar mass in the outskirts increases with redshift.}
\label{fig_stellar_fixed}
\end{figure}

Table \ref{tab_fluxes} presents the median, aperture-corrected flux densities measured in W1 and W2 stacks, with their uncertainties. We used {\tt CIGALE} \citep{burgarella_star_2005,noll_analysis_2009,boquien_cigale_2019} to calculate the integrated stellar masses in cylindrical annuli (i.e. without any modeling to deproject). Since {\tt CIGALE} does not allow single burst star formation histories, we assume a delayed $\mathrm{\tau}$-model with a characteristic time of 1 Myr and formation redshift of 3. 

Stellar masses for a fixed aperture are presented in the left and center panels of Figure \ref{fig_stellar_fixed}; the right panel shows a comparison between the mass in the outskirts and in the core. We also tested a formation redshift of 5 and found that at high redshift, the stellar masses are about 1.5 times higher than with $z_\mathrm{form}=3$.

The right panel of Figure \ref{fig_stellar_fixed} shows the ratio of these stellar masses, which removes some of the systematics associated with the assumed star formation history. The cluster outskirts were generally more populated at high redshift.

We note however that while subject to less systematics than the stellar mass measurements themselves, the ratios presented in the right panel of Figure \ref{fig_stellar_fixed} will be affected by spatial variations of the galaxy population within the cluster. In particular, there are multiple studies providing evidence that stellar populations are more evolved in the cores of clusters than in the outskirts \citep[e.g.][]{dressler_galaxy_1980,balogh_origin_2000,brodwin_era_2013,alberts_star_2016,nishizawa_first_2018,pintos-castro_evolution_2019,trudeau_massive_2024}. In addition, the fixed aperture used for the comparison might introduce a scale effect: the stellar mass ratios will be affected by the concentration of the clusters which might vary with redshift (see the explanation in Section \ref{sssec_ang_size}).

\bibliography{ref}{}
\bibliographystyle{aasjournalv7}



\end{document}